\documentclass[showpacs,preprintnumbers,amsmath,amssymb,twocolumn]{revtex4}
\usepackage{graphicx}
\usepackage{dcolumn}
\usepackage{bm}
\usepackage{amssymb}
\usepackage{array}
\usepackage{hhline}
\usepackage{longtable}
\begin{document}

\title{Distinct scalings for mean first-passage time of random walks on scale-free networks with the same degree sequence}

\author{Zhongzhi Zhang}
\email{zhangzz@fudan.edu.cn}

\author{Weilen Xie}

\author{Shuigeng Zhou}
\email{sgzhou@fudan.edu.cn}

\affiliation{School of Computer Science, Fudan University, Shanghai
200433, China} \affiliation{Shanghai Key Lab of Intelligent
Information Processing, Fudan University, Shanghai 200433, China}

\author{Mo Li}
\affiliation{Software School, Fudan University, Shanghai 200433,
China}

\author{Jihong Guan}
\email{jhguan@tongji.edu.cn}

\affiliation{Department of Computer Science and Technology, Tongji
University, 4800 Cao'an Road, Shanghai 201804, China}

\date{\today}

\begin{abstract}
In general, the power-law degree distribution has profound influence
on various dynamical processes defined on scale-free networks. In
this paper, we will show that power-law degree distribution alone
does not suffice to characterize the behavior of trapping problem on
scale-free networks, which is an integral major theme of interest
for random walks in the presence of an immobile perfect absorber. In
order to achieve this goal, we study random walks on a family of
one-parameter (denoted by $q$) scale-free networks with identical
degree sequence for the full range of parameter $q$, in which a trap
is located at a fixed site. We obtain analytically or numerically
the mean first-passage time (MFPT) for the trapping issue. In the
limit of large network order (number of nodes), for the whole class
of networks, the MFPT increases asymptotically as a power-law
function of network order with the exponent obviously different for
different parameter $q$, which suggests that power-law degree
distribution itself is not sufficient to characterize the scaling
behavior of MFPT for random walks, at least trapping problem,
performed on scale-free networks.
\end{abstract}




 \maketitle

\section{Introduction}

As a fundamental stochastic process, random walks have received
considerable attention from the scientific society, since they found
a wide range of distinct applications in various theoretical and
applied fields, such as physics, chemistry, biology, computer
science, among others~\cite{Sp1964,We1994,Hu1996}. Among a plethora
of interesting issues of random walks, trapping is an integral major
one, which plays an important role in an increasing number of
disciplines. The so-called trapping issue that was first introduced
in~\cite{Mo69}, is a random-walk problem, where a trap is positioned
at a fixed location, absorbing all particles that visit it. The
highly desirable quantity closed related to the trapping issue is
the first-passage time (FPT), also called trapping time (TT). The
FPT for a given site (node, vertex) is the time spent by the walker
starting from the site to hit the trap node for the first time. The
average of first-passage times over all nodes is referred to as the
mean first-passage time (MFPT), or mean trapping time (MTT), which
is frequently used to measure the efficiency of the trapping
problem.

One of the most important questions in the research of trapping is
determining its efficiency, namely, showing the dependence relation
of MFPT on the size of the system where the random walks are
performed. Previous studies have provided the answers to the
corresponding problems in some particular graphs with simple
structure, such as regular lattices~\cite{Mo69}, Sierpinski
fractals~\cite{KaBa02PRE,KaBa02IJBC}, T-fractal~\cite{Ag08}, and so
forth. However, recent empirical studies~\cite{AlBa02,DoMe02,Ne03}
uncovered that many (perhaps most) real networks are scale-free
characterized by a power-law degree distribution~\cite{BaAl99},
which cannot be described by above simple graphs. Thus, it appears
quite natural and important to explore the trapping issue on
scale-free networks. In recent
work~\cite{ZhQiZhXiGu09,ZhGuXiQiZh09,ZhZhXiChLiGu09}, we have shown
that scale-free property may substantially improve the efficiency of
the trapping problem: the MFPT behaves linearly or sublinearly with
the order (number of nodes) of the scale-free networks, which is in
sharp contrast to the superlinear scaling obtained for
above-mentioned simple graphs~\cite{Mo69,KaBa02PRE,KaBa02IJBC,Ag08}.
It was speculated that the high efficiency of trapping on scale-free
networks is attributed to their power-law property. Although
scale-free feature can strongly affect the various dynamics
occurring on networks, it was shown that the scale-free structure
itself does not suffice to characterize some dynamical processes on
networks, e.g., synchronization~\cite{AtBiJo06,HaSwSc06}, disease
spreading~\cite{EgKl02,ZhZhZoCh08}, and the like. Thus far, it is
still unknown whether the power-law degree distribution is
sufficient to characterize the behavior of trapping problem on
scale-free networks.

In this paper, we study the trapping problem on the a class of
scale-free networks with the same degree sequence, which are
dominated by a tunable parameter $q$~\cite{ZhZhZoChGu09}. We
determine separately the explicit formulas of the mean first-passage
time for the two limiting cases of $q=1$ and $q=0$. We show that in
both cases the MFPT increases as a power-law function of the network
order, with the exponent less than 1 for $q=1$ and equal to 1 for
$q=0$. We also study numerically the MFPT for the case of $0 < q <
1$, finding that it is also a power-law function of network order
with the exponent $\theta(q)$ depending on parameter $q$. We
demonstrate that in the full range of $0 \leq q \leq 1$, $\theta(q)$
is a decreasing function of $q$, which belongs to the interval
$\left [\frac{\ln 3}{\ln 4},1 \right]$. Our findings indicate that
the power-law degree distribution by itself is not sufficient to
characterize the trapping process taking place on scale-free
networks.

\section{The scale-free networks with identical degree sequence}

\begin{figure}[h]
\includegraphics[width=0.6\linewidth,trim=100 0 100 10]{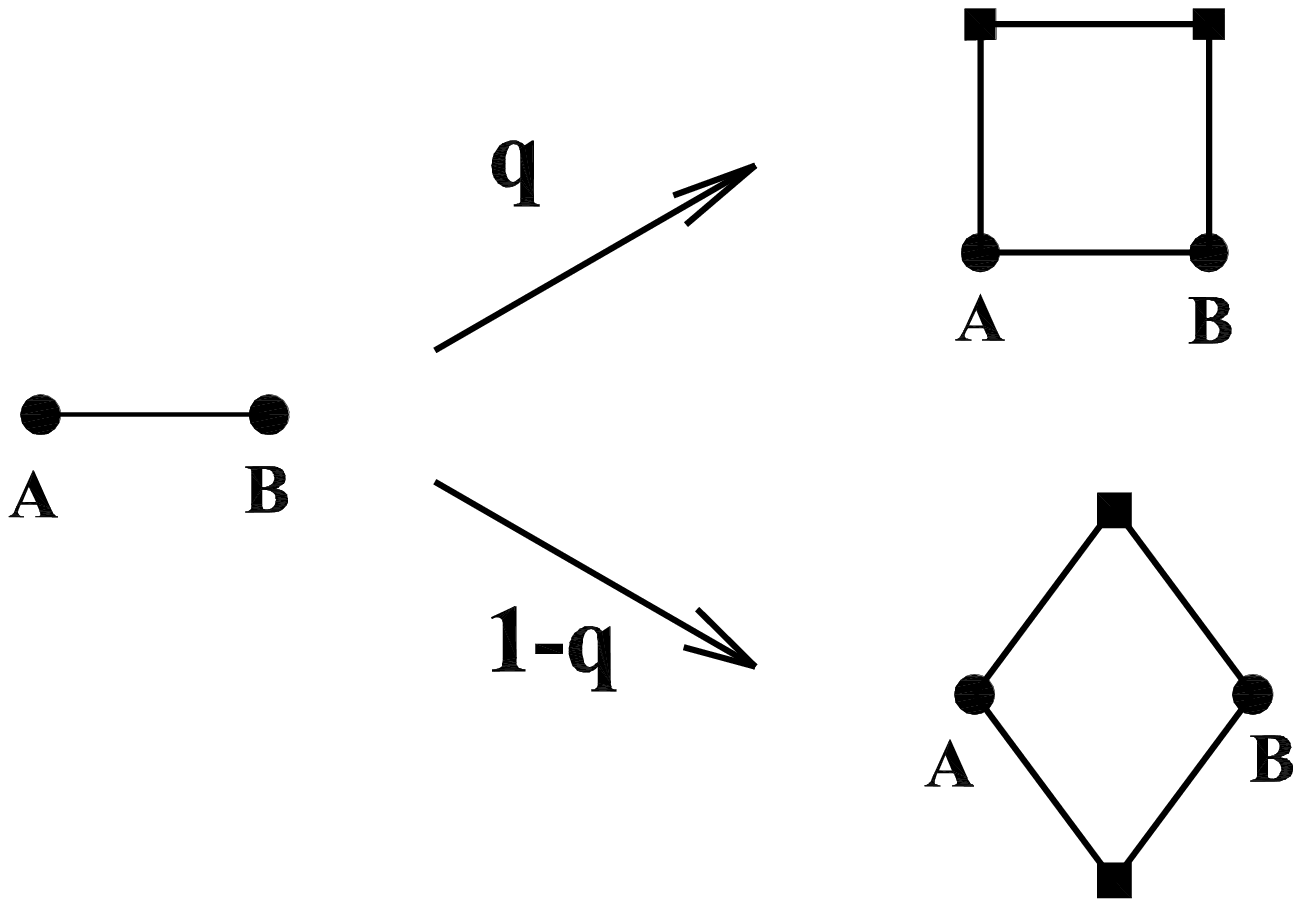}
\caption{Iterative method of the network construction. Each edge is
replaced by either of the connected clusters on the right-hand side
of arrows with a certain probability, where black squares represent
new nodes.}\label{fig1}
\end{figure}

\begin{figure}
\begin{center}
\includegraphics[width=1.00\linewidth,trim=100 10 100 15]{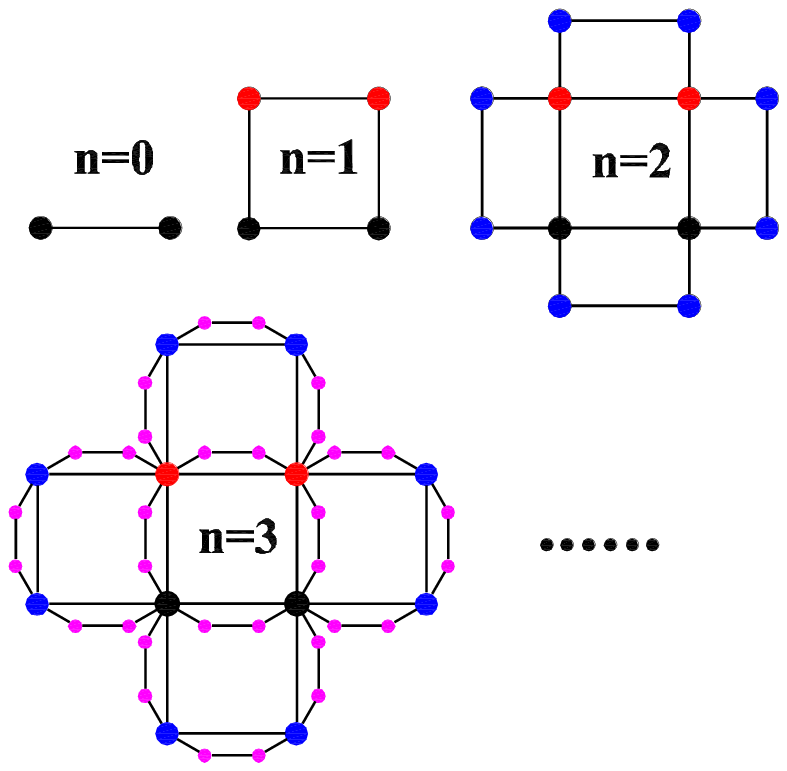}
\caption{(Color online) Illustration of the first several iterations
of the network for a particular case $q=1$.} \label{flower}
\end{center}
\end{figure}

\begin{figure}[h]
\centering
\includegraphics[width=0.85\linewidth,trim=80 0 120 20]{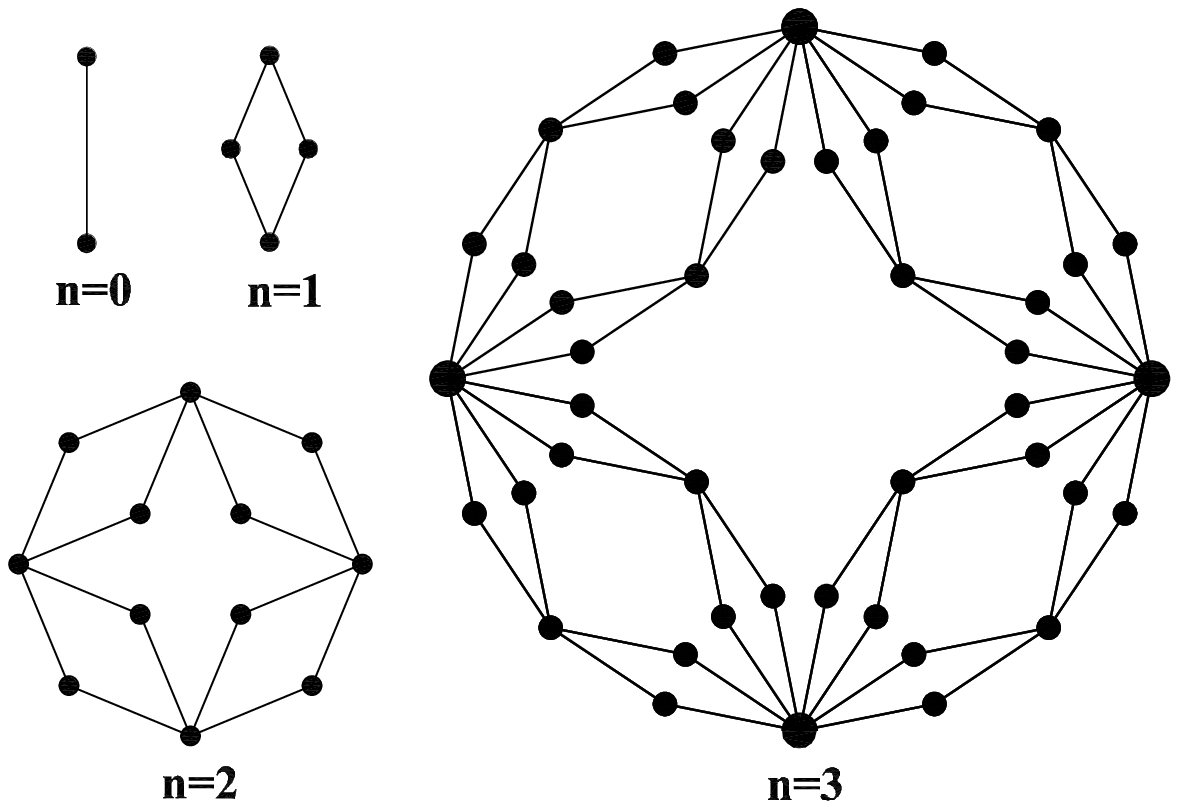}
\caption{Sketch of the iteration process of the network for the
limiting case of $q=0$.}\label{fractal}
\end{figure}

The networks in question are built iteratively~\cite{ZhZhZoChGu09},
see Fig.~\ref{fig1}. We represent by $H_{n}$ ($n \geq 0$) the
networks after $n$ iterations (the number of iterations is also
called generation hereafter). Then the networks are constructed as
follows. For $n=0$, the initial network $H_{0}$ consists of two
nodes connected to each other by an edge (a link). For $n \geq 1$,
$H_{n}$ is obtained from $H_{n-1}$. That is to say, to obtain
$H_{n}$, one can replace each link existing in $H_{n-1}$ either by a
connected cluster of links on the top right of Fig.~\ref{fig1} with
probability $q$, or by the connected cluster on the bottom right
with complementary probability $1-q$. Repeat the growth process $n$
times, with the graphs obtained in the limit $n \to \infty$. In
Figs.~\ref{flower} and~\ref{fractal}, we present the growing
processes of two special networks corresponding to $q=0$ and $q=1$,
respectively.

Let $L (n)$ be the number of nodes created at generation $n$, and
$E_n$ the total number of all edges present at generation $n$. By
construction, we have $E_n=4\,E_{n-1}$. Considering the initial
condition $E_0=2$, it leads to $E_n=4^n$. Since each existing edge
at a given generation will create two new nodes in at the next
generation, then, at each generation $n_i$ ($n_i\geq 1$) the number
of newly introduced nodes is $L(n_i)=2E_{n_i-1}=2\times4^{n_i-1}$.
Thus, at generation $n$ the network order is
\begin{equation}\label{Vn}
V_n=\sum_{n_i=0}^{n} L(n_i)=\frac{2}{3}(4^n+2)\,.
\end{equation}

Let $k_i(n)$ be the degree of a node $i$ at generation $n$, which
was created at generation $n_i$ ($n_i\geq 0$). Then,
\begin{equation}\label{ki01}
k_i(n)=2^{n-n_{i}+1}\,.
\end{equation}
From Eq.~(\ref{ki01}), it is obvious that after each new iteration
the degree of a node doubles, i.e.,
\begin{equation}\label{ki02}
k_i(n)=2\,k_i(n-1)\,.
\end{equation}

The networks considered exhibit some interesting topological
properties. Their nodes have same degree sequence (thus the same
degree distribution), independent of the value of parameter $q$.
Concretely, the networks have a power-law degree distribution with
the exponent $\gamma=3$~\cite{ZhZhZoChGu09}. On the other hand,
since there is no triangle in the whole class of the networks, the
clustering coefficient is zero. Although the degree distribution and
clustering coefficient do not depend on the parameter $q$, other
structural characteristics are closely related to $q$. For example,
for $q=1$, the network is reduced to the (2, 2)-flower introduced
in~\cite{RoHaAv07}. In this case, it is a small world, its average
path length (APL), defined as the mean of shortest distances between
all pairs of nodes, grows logarithmically with the network
order~\cite{ZhZhZoChGu09}; at the same time, it is a non-fractal
network~\cite{SoHaMa05,SoHaMa06}. While for $q=0$, it is exactly the
hierarchical lattice that was proposed by Berker and
Ostlund~\cite{BeOs79} and was extensively studied by many
authors~\cite{KaGr81,GrKa82,HiBe06,ZhZhZo07,RoAv07,ArBe09}. For this
case, the network is not small-world with the APL increasing as a
square power of the network order~\cite{HiBe06,ZhZhZo07}; moreover,
it is fractal with the fractal dimension $d_B=2$. When $q$ increases
from 0 to 1, the networks undergo a transition from fractal to
non-fractal scalings, and exhibit a crossover from `large' to small
worlds at the same time~\cite{ZhZhZoChGu09}; these similar phenomena
are also observed in a family of treelike
networks~\cite{ZhZhChGu08}.

The peculiar topological features make the networks unique within
the category of scale-free networks, since these particular
structures strongly affect the dynamical processes defined on the
networks. For instance, different thresholds of bond percolation
were recently observed in the networks, which implies that power-law
degree distribution alone does not suffice to characterize the
percolation threshold on scale-free networks under bond
percolation~\cite{ZhZhZoChGu09,RoAv07}. In what follows, we will
study random walks with a single immobile trap on the networks. We
will show that the degree distribution is not sufficient to
determine the scalings for MFPT of trapping process occurring on the
networks under consideration.

\section{Random walks with a fixed trap}

\begin{figure}
\begin{center}
\includegraphics[width=1.0\linewidth,trim=80 10 80 10]{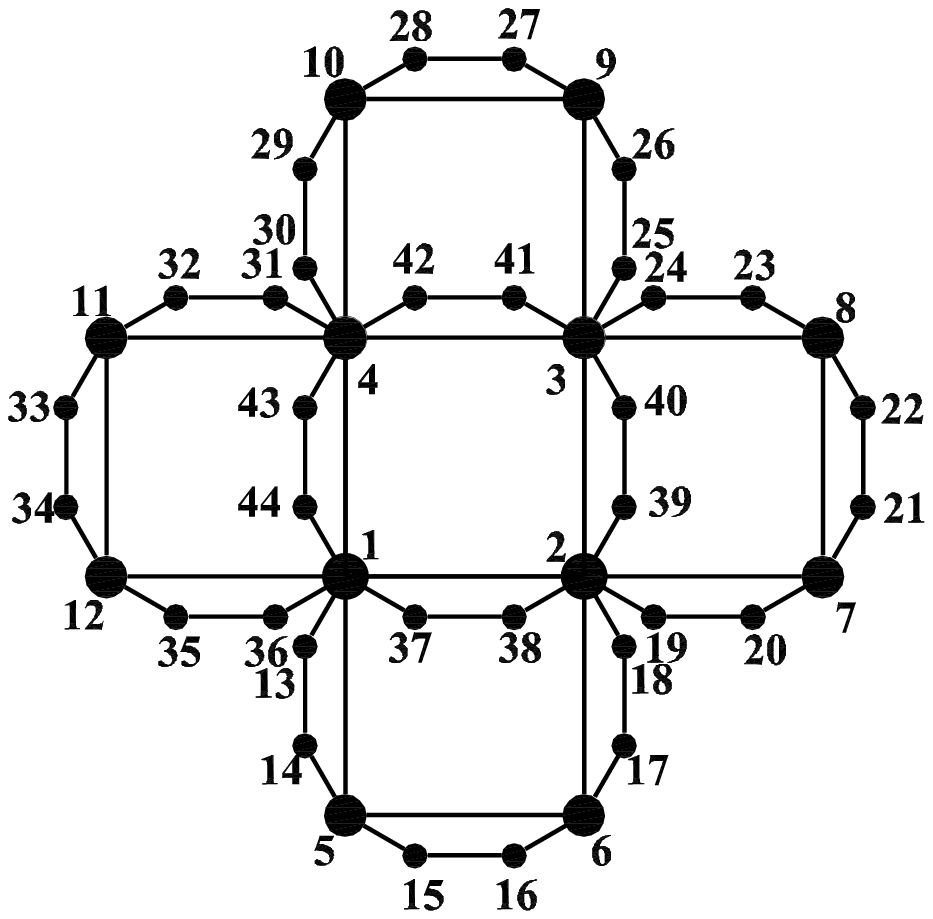}
\end{center}
\caption[kurzform]{\label{labeling01} labels of all nodes of ${H}_3$
in the case of $q=1$.}
\end{figure}

\begin{figure}
\begin{center}
\includegraphics[width=1.0\linewidth,trim=80 10 80 10]{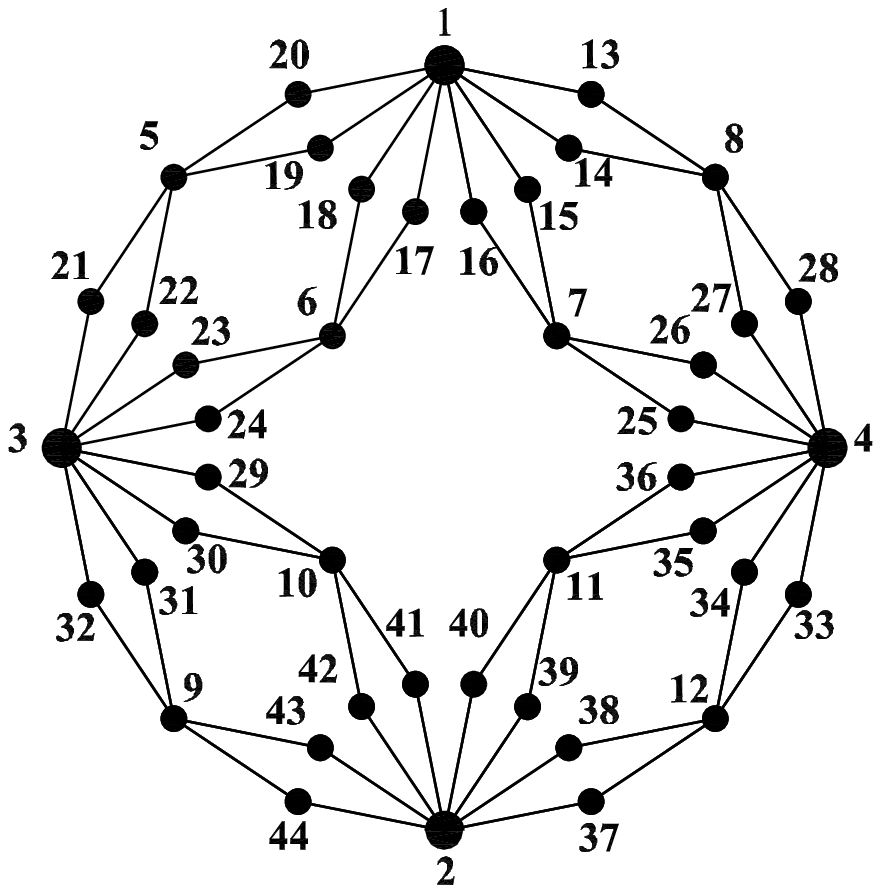}
\end{center}
\caption[kurzform]{\label{labeling02} labels of all nodes of ${H}_3$
for the particular case of $q=0$.}
\end{figure}

In this section, we study the so-called simple discrete-time random
walks of a particle on network $H_{n}$. At each time step, the
particle (walker) jumps from its current location to one of its
neighbors with equal probability. In particular, we focus on the
trapping problem, i.e., a special issue for random walks with a trap
positioned at a given node. To this end, we first we distinguish
different nodes in $H_{n}$, by labeling them in the following way.
The two nodes in $H_0$ have labels 1 and 2. For each new generation,
we only label the new nodes created at this generation, while we
keep the labels of all pre-existing nodes unchanged. In other words,
we label sequentially new nodes as $M+1, M+2,\ldots, M+\Delta M$,
where $M$ is the number of the old nodes and $\Delta M$ the number
of newly-created nodes. In this way, every node is labeled by a
unique integer, at generation $n$ all nodes are labeled from 1 to
$V_n=\frac{2}{3}(4^n+2)$. Figures~\ref{labeling01}
and~\ref{labeling02} show how the nodes are labeled for two special
cases of $q=1$ and $q=0$.

We place the trap at node 1, denoted by $i_T$. At each time step,
the particle, starting from any node except the trap $i_T$, moves
uniformly to any of its nearest neighbors.  It should be mentioned
that, due to the symmetry, the trap can be also situated at nodes 2,
3, or 4, which has not any effect on MFPT. The special selection we
made for the trap allows to address the issue conveniently.
Particularly, this makes it possible to analytically compute the
MFPT for the two deterministic networks corresponding to $q=1$ and
$q=0$ (details will be discussed below), because of their special
structures and the convenience of identifying the trap $i_T$ since
the first generation.

As mentioned above, one of the most important quantity
characterizing such a trapping problem is the FPT defined as the
expected time a walker takes, starting from a source node, to first
reach the trap node. The significance firstly originates from the
fact that the first encounter properties are relevant to those in a
plethora of real situations~\cite{CoBeTeVoKl07}, including
transport, disease spreading, target search, and so on. On the other
hand, many other quantities can be expressed in terms of FPTs, and
more information about the dynamics of random walks can be extracted
from the analysis of FPTs~\cite{Re01}. Finally, the average of
first-passage times, i.e., mean first-passage time (MFPT), measures
the efficiency of the trapping process: the smaller the MFPT, the
higher the efficiency, and vice versa. In the following, we will
determine the exact solutions to MFPT for some limiting cases, as
well as the dependence relation of MFPT on the network order.

Let $T_i^{(n)}$ be the FPT for a walker initially placed at node $i$
to first reach the trap $i_T$ in $H_n$. This quantity can be
expressed in terms of mean residence time
(MRT)~\cite{BaKl98,ZhZhXiChLiGu09}, which is defined to be the mean
time that a random walker spends at a given node prior to being
absorbed by the trap. Actually, the MRT is the mean number of
visitations of a given node by the walker before trapping occurs.

It is known that the trapping problem studied can be described by a
Markov chain~\cite{KeSn76}, whose fundamental matrix is the inverse
of matrix $\mathbf{B}_n$ that is a variant of the normalized
Laplacian matrix~\cite{Ch97} $\mathbf{L}_n$ for $H_n$, and can be
obtained from $\mathbf{L}_n$ with all entries in the first row and
column (corresponding to the trap node) setting to zeros. The entry
$(b_n^{-1})_{ij}$ of the fundamental matrix $(\mathbf{B}_n)^{-1}$
expresses the mean number of visitations of node $j$ by the
particle, starting from node $i$, before it is eventually trapped.
Thus, we have
\begin{equation}\label{MFPT4}
T_i^{(n)}=\sum_{j=2}^{V_n}(b_n^{-1})_{ij}\,.
\end{equation}
Then, the mean first-passage time, $\langle T \rangle_n$, which is
the average of $T_i^{(n)}$ over all initial nodes distributed
uniformly over nodes in $H_n$ other than the trap, is given by
\begin{equation}\label{MFPT5}
 \langle T
\rangle_n=\frac{1}{V_n-1}\sum_{i=2}^{V_n}
T_i^{(n)}=\frac{1}{V_n-1}\sum_{i=2}^{V_n}\sum_{j=2}^{V_n}(b_n^{-1})_{ij}\,.
\end{equation}

Equation~(\ref{MFPT5}) shows that the problem of determining
$\langle T \rangle_n$ is reduced to computing the sum of all
elements of the fundamental matrix $(\mathbf{B}_n)^{-1}$. Although
the expression of Eq.~(\ref{MFPT5}) seems compact, the complexity of
inverting $\mathbf{L}_n$ is \emph{O}($V_n^3$). Since the network
order increases exponentially with $n$, Eq.~(\ref{MFPT5}) becomes
intractable for large $n$. Thus, restricted by time and computer
memory, one can obtain $\langle T \rangle_n$ through direct
calculation from Eq.~(\ref{MFPT5}) only for the first iterations. It
would be satisfactory if good alternative computation methods could
be proposed to get around this problem. What is encouraging is that
the particular construction of the networks and the special choice
of the trap location allow to calculate analytically MFPT to obtain
a closed-form formula, at least for the two special cases of $q=1$
and $q=0$. The computation details will be provided in the following
text.


\subsection{Case of $q = 1$}

We first establish the scaling relation governing the evolution for
$T_i^{(n)}$ with generation $n$. In Table~\ref{tab:AMTA2q1}, we list
the numerical values of $T_i^{(n)}$ for some nodes up to $n=6$. From
the numerical values, we can observe that for a given node $i$, the
relation $T_i^{(n+1)}=3\,T_i^{(n)}$ holds. That is to say, upon
growth of the network from generation $n$ to generation $n+1$, the
trapping time to first arrive at the trap increases by a factor $3$.
This is a basic characteristic of random walks on $H_n$ when $q =
1$, which can be established from the arguments
below~\cite{HaBe87,KaRe89,Bobe05}.

\begin{table*}
\caption{Numerical results of the trapping time $T_i^{(n)}$ for a
random walker starting from node $i$ on the network $H_n$ for
various $n$ in the case of $q=1$. All the values are obtained
through the direct calculation from Eq.~(\ref{MFPT4}).}
\label{tab:AMTA2q1}
\begin{center}
\begin{tabular}{l|cccccccccccccccccccccc}
\hline \hline $n \backslash i$  &2,3& 4&5,6&7,8 \quad & 9,10 \quad
    & 11,12 \quad &13,14 \quad&15,16\quad &17,18\quad &19,20 \quad
    & 21,22 \quad& 23,24 \quad& 25,26 \quad& 27,28 \quad& 29-32 \quad& 33-36 \quad& 37-40\quad
    & 41-44 \\
\hline\hline
            1 & $3$ & $4$       \\
            2 & $9$ & $12$ & $5$ & $8$ & $12$ & $13$  \\
            3 & $27$ & $36$ & $15$ & $24$ & $36$ & $39$  & $7$      & $12$ & $20$   & $23$
                & $27$ & $28$ & $11$ & $20$ & $32$ & $35$ & $39$ & $40$ \\
            4 & $81$ & $108$ & $45$ & $72$ & $108$ & $117$  & $21$  &   $36$&$60$   & $69$
                & $81$ & $84$ & $33$ & $60$ & $96$        & $105$& $117$ & $120$ \\
            5 & $243$ & $324$ & $135$ & $216$ & $324$ & $351$  & $63$&  $108$&$180$ & $207$
                & $243$ & $252$ & $99$ & $180$ & $288$    & $315$& $351$ & $360$ \\
            6 & $729$ & $972$ & $405$ & $648$ & $972$ & $1053$  & $189$&$324$&$540$ & $621$
                & $729$ & $756$ & $297$ & $540$ & $864$   & $945$& $1053$& $1080$ \\
\hline \hline
\end{tabular}
\end{center}
\end{table*}

Consider an arbitrary node $i$ in $H_n$ of the $q=1$ case, after $n$
generation evolution. From Eq.~(\ref{ki02}), we know that upon
growth of the network to generation $n+1$, the degree, $k_i$, of
node $i$ doubles, namely, it increases from $k_i$ to $2\,k_i$. Among
these $2\,k_i$ neighbors, one half are old neighbors, while the
other half are new nodes created at generation $n+1$, each of which
has two connections, attached to node $i$ and another simultaneously
emerging new node. We now examine the standard random walk in
$H_{n+1}$: Let $X$ be the FPT for a particle going from node $i$ to
any of its $k_i$ old neighbors; let $Y$ be the FPT for going from
any of the $k_i$ new neighbors of $i$ to one of the $k_i$ old
neighbors; and let $Z$ represent the FPT for starting from any of
new neighbors (added to the network at generation $n+1$) of an old
neighbor of $i$ to this old neighbor. Then we can establish the
following backward equations:
\begin{eqnarray}\label{MFPT6}
\left\{
\begin{array}{ccc}
X&=&\frac{1}{2} + \frac{1}{2}(1+Y)\,,\\
Y&=&\frac{1}{2}(1+X) + \frac{1}{2}(1+Z)\,,\\
Z&=&\frac{1}{2} + \frac{1}{2}(1+Y)\,.
 \end{array}
 \right.
\end{eqnarray}

Equation~(\ref{MFPT6}) has a solution $X=3$. Thus, upon the growth
of the network from generation $n$ to generation $n+1$, the
first-passage time from any node $i$ to any node $j$ (both $i$ and
$j$ belong to $H_n$) increases by a factor of 3. That is to say,
$T_i^{(n+1)}=3\,T_i^{(n)}$, which will be useful for the derivation
of the exact formula for the MFPT below.

After obtaining the scaling of first-passage time for old nodes, we
now derive the analytical rigorous expression for the MFPT $\langle
T \rangle_n$. Before proceeding further, we first introduce the
notations that will be used in the rest of this section. Let
$\Delta_n$ denote the set of nodes in $H_n$, and let
$\overline{\Delta}_n$ stand for the set of those nodes entering the
network at generation $n$. For the convenience of computation, we
define the following quantities for $1\leq m \leq n$:
\begin{equation}
    T_{m, {\rm tot}}^{(n)} = \sum_{i \in \Delta_m} T_i^{(n)},
\end{equation}
and
\begin{equation}
    \overline{T}_{m, {\rm tot}}^{(n)} = \sum_{i \in \overline{\Delta}_m} T_i^{(n)}.
\end{equation}

By definition, it follows that $\Delta_n = \overline{\Delta}_n \cup
\Delta_{n-1}$. Thus, we have
\begin{equation}\label{eq:Ttotq1}
T_{n, {\rm tot}}^{(n)} = T_{n - 1, {\rm tot}}^{(n)} +
\overline{T}_{n, {\rm tot}}^{(n)}
        = 3\,T_{n - 1, {\rm tot}}^{(n-1)} + \overline{T}_{n, {\rm
        tot}}^{(n)}\,,
\end{equation}
where the relation of $T_i^{(n+1)}=3\,T_i^{(n)}$ has been made use
of. Hence, in order to  determine $\overline{T}_{n, {\rm
tot}}^{(n)}$, we should first find the quantity $\overline{T}_{n,
{\rm tot}}^{(n)}$ that can be obtained as follows.

\begin{figure}[h]
\begin{center}
\includegraphics[width=.5\linewidth,trim=50 20 50 0]{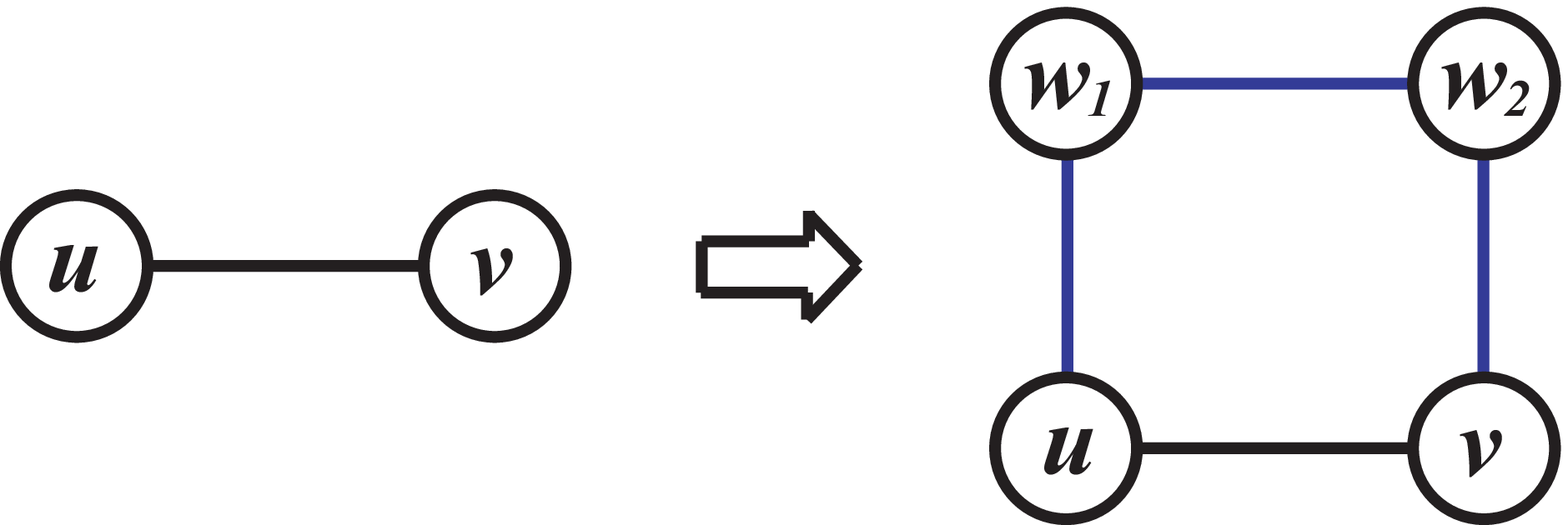}
\caption{(Color online) Illustration showing the relation of the
mean transmit times for two new nodes and two old nodes connected by
an edge generating the new nodes.}\label{iter}
\end{center}
\end{figure}

By construction, at a given generation, for each edge connecting two
nodes $u$ and $v$ (see Fig.~\ref{iter}), it will generate two new
nodes (say $w_1$ and $w_2$) in the next generation, and the mean
transmit times for the two new nodes obey the following relations:
\begin{eqnarray}\label{MFPT6q1}
\left\{
\begin{array}{ccc}
T(w_1) &=& \frac{1}{2}(1+T(w_2))+\frac{1}{2}\left(1+T(u)\right)\,,\\
T(w_2) &=& \frac{1}{2}(1+T(w_1))+\frac{1}{2}\left(1+T(v)\right)\,.
 \end{array}
 \right.
\end{eqnarray}
Hence, we have
\begin{equation}\label{eq:oldnewq1}
        T(w_1)+T(w_2) = 4 + T(u) + T(v)\,.
\end{equation}
Summing Eq.~(\ref{eq:oldnewq1}) over all the $E_n$ old edges
pre-existing at the generation $n$, we obtain
\begin{eqnarray}
    \overline{T}_{n+1, {\rm tot}}^{(n+1)} &=& 4\,E_n +
    \sum_{i \in \Delta_{n}}\left(k_i(n)\times T_i^{(n)}\right) \nonumber\\
    &=& 4^{n+1} + 2\,\overline{T}_{n, {\rm tot}}^{(n+1)}
    +2^2\,\overline{T}_{n-1, {\rm tot}}^{(n+1)}  + \ldots \nonumber\\ &\quad& + 2^n\,\overline{T}_{1, {\rm tot}}^{(n+1)}+ 2^n\,\overline{T}_{0, {\rm
    tot}}^{(n+1)}\,.
    \label{eq:rec1q1}
\end{eqnarray}

For example, in $H_2$ (see Fig.~\ref{labeling01}), $\overline{T}_{2,
{\rm tot}}^{(2)}$ can be expressed as
\begin{eqnarray}
    \overline{T}_{2, {\rm tot}}^{(2)} &=& \left(T_{5}^{(2)} + T_{6}^{(2)}\right) + \left(T_{7}^{(2)}+ T_{8}^{(2)}\right)    \nonumber\\
    &\quad& {} + \left(T_{9}^{(2)} + T_{10}^{(2)}\right) + \left(T_{11}^{(2)} + T_{12}^{(2)}\right) \nonumber\\
    &=& \left(4+T_{1}^{(2)}+T_{2}^{(2)}\right) + \left(4+T_{2}^{(2)}+T_{3}^{(2)}\right) \nonumber\\
    &\quad& + \left(4+T_{3}^{(2)}+T_{4}^{(2)}\right) + \left(4+T_{4}^{(2)}+T_{1}^{(2)}\right) \nonumber\\
    &=& 4\,E_{1} + 2\left(T_1^{(2)} + T_2^{(2)} + T_3^{(2)} + T_4^{(2)}\right)  \nonumber\\
    &=& 16 + 2\,\overline{T}_{1, {\rm tot}}^{(2)}+ 2\,\overline{T}_{0, {\rm
    tot}}^{(2)}\,.
 \end{eqnarray}
Again, for instance, in $H_3$ (see Fig.~\ref{labeling01}),
$\overline{T}_{3, {\rm tot}}^{(3)}$ can be written as
\begin{equation}
\overline{T}_{3, {\rm tot}}^{(3)} = 64 + 2\,\overline{T}_{2, {\rm
tot}}^{(3)} + 4\,\overline{T}_{1, {\rm
tot}}^{(3)}+4\,\overline{T}_{0, {\rm tot}}^{(3)}\,.
\end{equation}

Now, we can determine $\overline{T}_{n, {\rm tot}}^{(n)}$ through a
recurrence relation, which can be obtained easily. From
Eq.~(\ref{eq:rec1q1}), it is not difficult to write out
$\overline{T}_{n+2, {\rm tot}}^{(n+2)}$ as
\begin{eqnarray}
\overline{T}_{n+2, {\rm tot}}^{(n+2)} = 4^{n+2}&+&
2\,\overline{T}_{n+1, {\rm tot}}^{(n+2)} + 2^2\,\overline{T}_{n,
{\rm tot}}^{(n+2)} + \ldots \nonumber\\
    &+& 2^{n+1}\,\overline{T}_{1, {\rm tot}}^{(n+2)}+ 2^{n+1}\,\overline{T}_{0, {\rm tot}}^{(n+2)}\,.  \label{eq:rec2q1}
\end{eqnarray}

Equation~(\ref{eq:rec2q1}) minus Eq.~(\ref{eq:rec1q1}) times 6 and
applying the relation of $T_i^{(n+2)}=3\,T_i^{(n+1)}$, one gets the
following recurrence relation
\begin{equation} \label{eq:T1q1}
    \overline{T}_{n+2, {\rm tot}}^{(n+2)} = 12\,\overline{T}_{n+1, {\rm tot}}^{(n+1)} - 2 \times 4^{n+1}\,.
\end{equation}
Using $\overline{T}_{1, {\rm tot}}^{(1)} = 7$, Eq.~(\ref{eq:T1q1})
is solved inductively
\begin{equation}    \label{eq:TX1q1}
    \overline{T}_{n, {\rm tot}}^{(n)} = 4^{n-1} + 6 \times
    12^{n-1}\,.
\end{equation}

Inserting Eq.~(\ref{eq:TX1q1}) into Eq.~(\ref{eq:Ttotq1}) leads to
\begin{equation}
    T_{n, {\rm tot}}^{(n)} = 3\,T_{n-1, {\rm tot}}^{(n-1)}+
    4^{n-1} + 6 \times 12^{n-1}\,.  \label{eq:T2q1}
\end{equation}
Considering the initial condition $T_{1, {\rm tot}}^{(1)} = 10$,
Eq.~(\ref{eq:T2q1}) is resolved by induction to obtain
\begin{equation}\label{eq:T2q12}
    T_{n, {\rm tot}}^{(n)} = \frac{2}{3} \times 12^n + 4^n - 2
    \times 3^{n-1}\,.
\end{equation}
Substituting Eq.~(\ref{eq:T2q12}) into Eq.~(\ref{MFPT5}), we obtain
the closed-form expression for the MFPT for the trapping problem on
$H_n$ of the $q=1$ case as follows:
\begin{equation}\label{MFPT7}
 \langle T
\rangle_n=\frac{1}{V_n-1}T_{n, {\rm tot}}^{(n)}=\frac{1}{2\times
4^n+1}(2\times 12^n + 3\times4^n - 2
    \times 3^{n})\,.
\end{equation}

Below we will show how to express $\langle T \rangle_n$ in terms of
network order $V_n$, with the aim of obtaining the relation between
these two quantities. Recalling Eq.~(\ref{Vn}), we have
$4^{n}=\frac{3}{2}V_n-2$ and $n=\log_4\big(\frac{3}{2}V_n-2\big)$.
Thus, Eq.~(\ref{MFPT7}) can be rewritten as
\begin{equation}\label{MFPT20}
\langle T
\rangle_n=\frac{V_n-2}{V_n-1}\left(\frac{3}{2}V_n-2\right)^{\frac{\ln
3}{\ln 4}}+\frac{3V_n-4}{2(V_n-1)}.
\end{equation}
For large network, i.e., $V_n\rightarrow \infty$,
\begin{equation}\label{MFPT21}
\langle T \rangle_n \sim (V_n)^{\ln 3/ \ln 4}\,,
\end{equation}
with the exponent less than 1. Thus, in large network the MFPT grows
sublinearly with network order.

\begin{table*}
\caption{The trapping time $T_i^{(n)}$ for a random walker starting
from node $i$ on the network $H_n$ for various $n$ in the case of
$q=0$. All the values are calculated straightforwardly from
Eq.~(\ref{MFPT4}).} \label{tab:AMTA2q0}
\begin{center}
\begin{tabular}{l|cccccccccc}
\hline \hline  $n \backslash i$  &2 \quad & \quad
3,4 \quad\quad &5-8 \quad &\quad 9-12 \quad & \quad 13-20 \quad &\quad 21-28 \quad &\quad 29-36 \quad &\quad 37-44 \\
\hline \hline
            1 & $4$ & $3$       \\
            2 & $16$ & $12$ & $7$ & $15$   \\
            3 & $64$ & $48$ & $28$ & $60$ & $15$ & $39$         & $55$ & $63$  \\
            4 & $256$ & $192$ & $112$ & $240$ & $60$ & $156$    & $220$ & $252$  \\
            5 & $1024$ & $768$ & $448$ & $960$ & $240$ & $624$  & $880$ & $1008$  \\
            6 & $4096$ & $3072$ & $1792$ & $3840$ & $960$ & $2496$ & $3520$ & $4032$  \\
\hline \hline
\end{tabular}
\end{center}
\end{table*}

\subsection{Case of $q = 0$}

Analogous to the case of $q=1$, before deriving the general formula
for $\langle T\rangle_n$ for the limiting case of $q=0$, we first
establish the scaling relation dominating $T_i^{(n)}$ evolving with
generation $n$. To attain this goal, we examine the numerical values
of $T_i^{(n)}$ for some nodes up to $n=6$, which can be obtained
straightforwardly via equation~(\ref{MFPT4}). From the numerical
results listed in Table~\ref{tab:AMTA2q0}, one can easily observe
that for a given node $i$, its MFPT changes with the generation as
$T_i^{(n+1)}=4\,T_i^{(n)}$, which can be supported by the following
argument.

Consider a node $i$ in the $n$th generation of network $H_n$ for a
particular case of $q=0$. In the generation $n+1$, its degree $k_i$
doubles by growing from $k_i$ to $2\,k_i$. Moreover, different from
that of the $q=1$ case, all the $2\,k_i$ neighbors of node $i$ are
new nodes created at generation $n+1$. We now examine the random
walks taking place in $H_{n+1}$: Let $X$ be the FPT originating at
node $i$ to any of its $k_i$ old neighbors, i.e., those nodes
directly connected to $i$ at iteration $n$; and let $Y$ denote FPT
for going from any of the $2\,k_i$ new neighbors of $i$ to one of
its $k_i$ old neighbors. Then the following relations hold:
\begin{eqnarray}\label{MFPT6q0}
\left\{
\begin{array}{ccc}
X&=&1+Y\,   \\
Y&=&\frac{1}{2} + \frac{1}{2}(1+X)\,.
 \end{array}
 \right.
\end{eqnarray}
Equation~(\ref{MFPT6q0}) has a solution $X=4$ found by eliminating
$Y$, which means that for any pair of nodes $i$ and $j$ in $H_n$,
the FPT from $i$ to $j$ increases by a factor of 4 during the growth
of the network from generation $n$ to generation $n+1$. The relation
$T_i^{(n+1)}=4\,T_i^{(n)}$ is a basic feature for random walks on
the network of the $q=0$ case, which will be applied to the
derivation of the exact formula for $\langle T\rangle_n$.

Having obtaining the evolution relation of trapping time for old
nodes when the network grows, we continue to derive the analytical
rigorous expression for the MFPT. In what follows, we will use the
same notations as those for the $q=1$ case defined above. Similar to
the $q=1$ case, it is easy to get the following equation:
\begin{equation}\label{eq:Ttotq0}
T_{n, {\rm tot}}^{(n)} = T_{n - 1, {\rm tot}}^{(n)} +
\overline{T}_{n, {\rm tot}}^{(n)}
        = 4\,T_{n - 1, {\rm tot}}^{(n-1)} + \overline{T}_{n, {\rm
        tot}}^{(n)}\,.
\end{equation}

\begin{figure}[h]
\begin{center}
\includegraphics[width=.5\linewidth,trim=50 20 50 0]{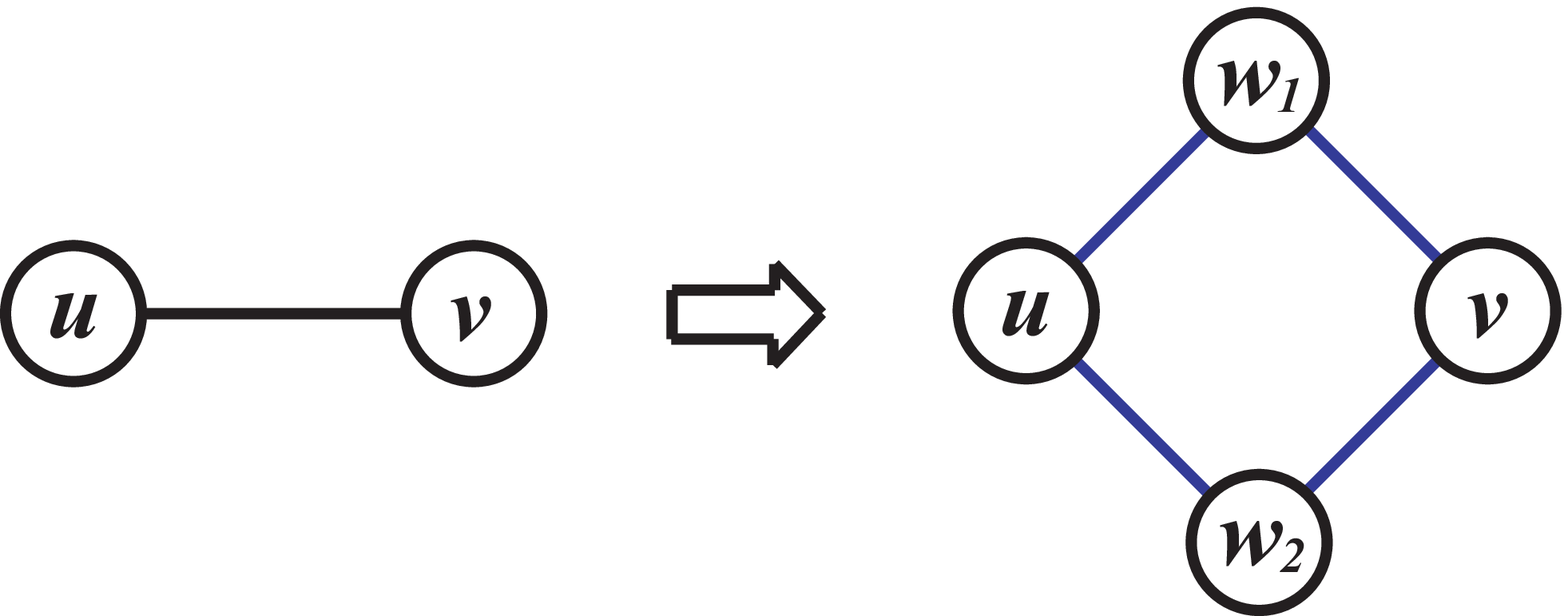}
\caption{(Color online) Illustration showing the relation of the
first passage times for two new nodes and two old nodes connected at
last generation by an edge creating the new nodes.}\label{iter1}
\end{center}
\end{figure}

Therefore, to determine $T_{n, {\rm tot}}^{(n)}$, we need to find
$\overline{T}_{n, {\rm tot}}^{(n)}$ first, which can be obtained as
follows. Notice that for any given edge attaching two nodes $u$ and
$v$ (see Fig.~\ref{iter1}) in $H_n$, it will generate two new nodes
($w_1$ and $w_2$) in $H_{n+1}$, and the FPTs for the two new nodes
are equal to each other obeying the following equation:
\begin{equation}\label{MFPTq0}
 T(w_1)= T(w_2)=1+\frac{1}{2}\left(T(u)+T(v)\right)\,,
\end{equation}
which yields to
\begin{equation}\label{eq:oldnewq0}
        T(w_1)+T(w_2) = 2 + T(u) + T(v)\,.
\end{equation}

Summing Eq.~(\ref{eq:oldnewq0}) over all $E_n$ old edges belonging
to $H_n$, we have
\begin{eqnarray}
    \overline{T}_{n+1, {\rm tot}}^{(n+1)} &=& 2\,E_n +
    \sum_{i \in \Delta_{n}}\left(k_i(n)\times T_i^{(n)}\right) \nonumber\\
    &=&  2\times 4^{n} + 2\,\overline{T}_{n, {\rm tot}}^{(n+1)}
    +2^2\,\overline{T}_{n-1, {\rm tot}}^{(n+1)}  + \ldots \nonumber\\ &\quad& + 2^n\,\overline{T}_{1, {\rm tot}}^{(n+1)}+ 2^n\,\overline{T}_{0, {\rm
    tot}}^{(n+1)}\,,
    \label{eq:rec1q0}
\end{eqnarray}
from which we can derive the following recursive relation:
\begin{equation} \label{eq:T1q0}
    \overline{T}_{n+2, {\rm tot}}^{(n+2)} = 16\,\overline{T}_{n+1, {\rm tot}}^{(n+1)} - 2 \times
    4^{n+1}\,.
\end{equation}
Using the initial condition $\overline{T}_{1, {\rm tot}}^{(1)} = 6$,
Eq.~(\ref{eq:T1q0}) is solved inductively to get
\begin{equation} \label{eq:TX1q0}
    \overline{T}_{n, {\rm tot}}^{(n)} = \frac{4^{n}}{6} +
    \frac{4^{2n}}{3}\,.
\end{equation}

Plugging Eq.~(\ref{eq:TX1q0}) into Eq.~(\ref{eq:Ttotq0}), we have
\begin{equation}    \label{eq:T2q0}
    T_{n, {\rm tot}}^{(n)} = 4\,T_{n-1,{\rm
    tot}}^{(n-1)}+\frac{4^{n}}{6}+\frac{4^{2n}}{3}\,.
\end{equation}
Combining with the initial condition $T_{1, {\rm tot}}^{(1)} = 10$,
one can solve Eq.~(\ref{eq:T2q0}) by induction to obtain
\begin{equation}\label{eq:ToTq0}
    T_{n, {\rm tot}}^{(n)} = \frac{4^{n}}{18}(8\times 4^{n}+3n+10)\,.
\end{equation}
Inserting Eq.~(\ref{eq:ToTq0}) into Eq.~(\ref{MFPT5}), we obtain the
rigorous solution for the MFPT for the trapping issue performed on
$H_n$ of the $q=0$ case:
\begin{equation}\label{MFPT8}
 \langle T
\rangle_n=\frac{1}{V_n-1}T_{n, {\rm tot}}^{(n)}=\frac{4^n}{6(2\times
4^n+1)}(8\times 4^{n}+3n+10)\,.
\end{equation}
As in the case of $q=1$, we can recast $\langle T \rangle_n$ as a
function of the network order:
\begin{equation}\label{MFPT9}
 \langle T
\rangle_n=\frac{3V_n-4}{36(V_n-1)}\left[12V_n+ \frac{3\,\ln
\left(\frac{3}{2}V_n-2\right)}{2\,\ln 2}-6\right],
\end{equation}
from which it is easy to see that for large network (i.e.,
$V_n\rightarrow \infty$), we have the following expression:
\begin{equation}\label{MFPT21}
\langle T \rangle_n \approx V_n\, .
\end{equation}
Thus, the MFPT grows linearly with increasing order of network,
which is in sharp contrast to the sublinear scaling for the $q=1$
case shown above.

\begin{figure}
\begin{center}
\includegraphics[width=0.85\linewidth,trim=30 30 20 0]{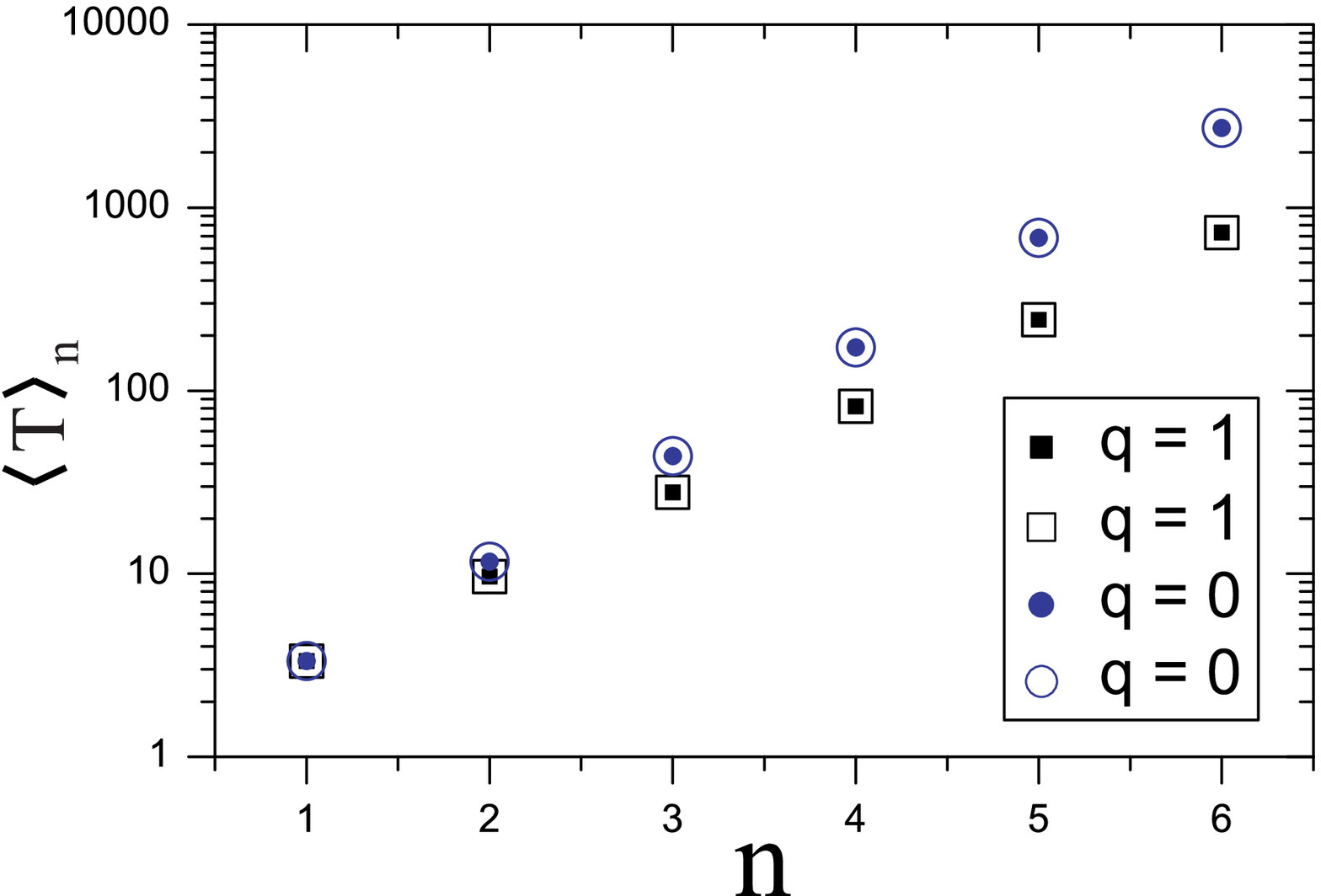}
\end{center}
\caption[kurzform]{\label{Time01} (Color online) Mean first-passage
time $\langle T \rangle_n$ as a function of the iteration $n$ on a
semilogarithmic scale for two case of $q=1$ and $q=0$. The filled
symbols are the numerical results obtained by direct calculation
from Eq.~(\ref{MFPT5}), while the empty symbols correspond to the
exact values from Eqs.~(\ref{MFPT7}) and~(\ref{MFPT8}). The
analytical and numerical values are consistent with each other.}
\end{figure}

In order to confirm the analytical expressions provided by
Eqs.~(\ref{MFPT7}) and~(\ref{MFPT8}), we have compared the exact
solutions for the MFPT with numerical values given by
Eq.~(\ref{MFPT5}), see Fig.~\ref{Time01}. For all $1 \leq n \leq 6$,
the analytical values obtained from Eqs.~(\ref{MFPT7})
and~(\ref{MFPT8}) show complete agreement with their corresponding
numerical results. This agreement is an independent test of our
theoretical formulas.

\subsection{Case of $0 < q < 1$}

We have obtained the explicit expressions for MFPT of random walks
with a trap on the networks for two limiting cases of $q=1$ and
$q=0$, and shown that for the corresponding cases the MFPT grows
sublinearly or linearly with the network order. But for the case of
$0 < q < 1$, there are some difficulties in obtaining a closed
formula for $\langle T\rangle_n$ as for the two special cases of
$q=1$ and $q=0$, since for $q=1$ and $q=0$, the networks are
deterministic and self-similar, which allows one derive the the
analytic solutions for $\langle T\rangle_n$; while for $0 < q < 1$,
the networks are stochastic, which makes it impossible to write a
recursive relation for the evolution of the first-passage time.

\begin{figure}
\begin{center}
\includegraphics[width=0.85\linewidth,trim=30 30 20 0]{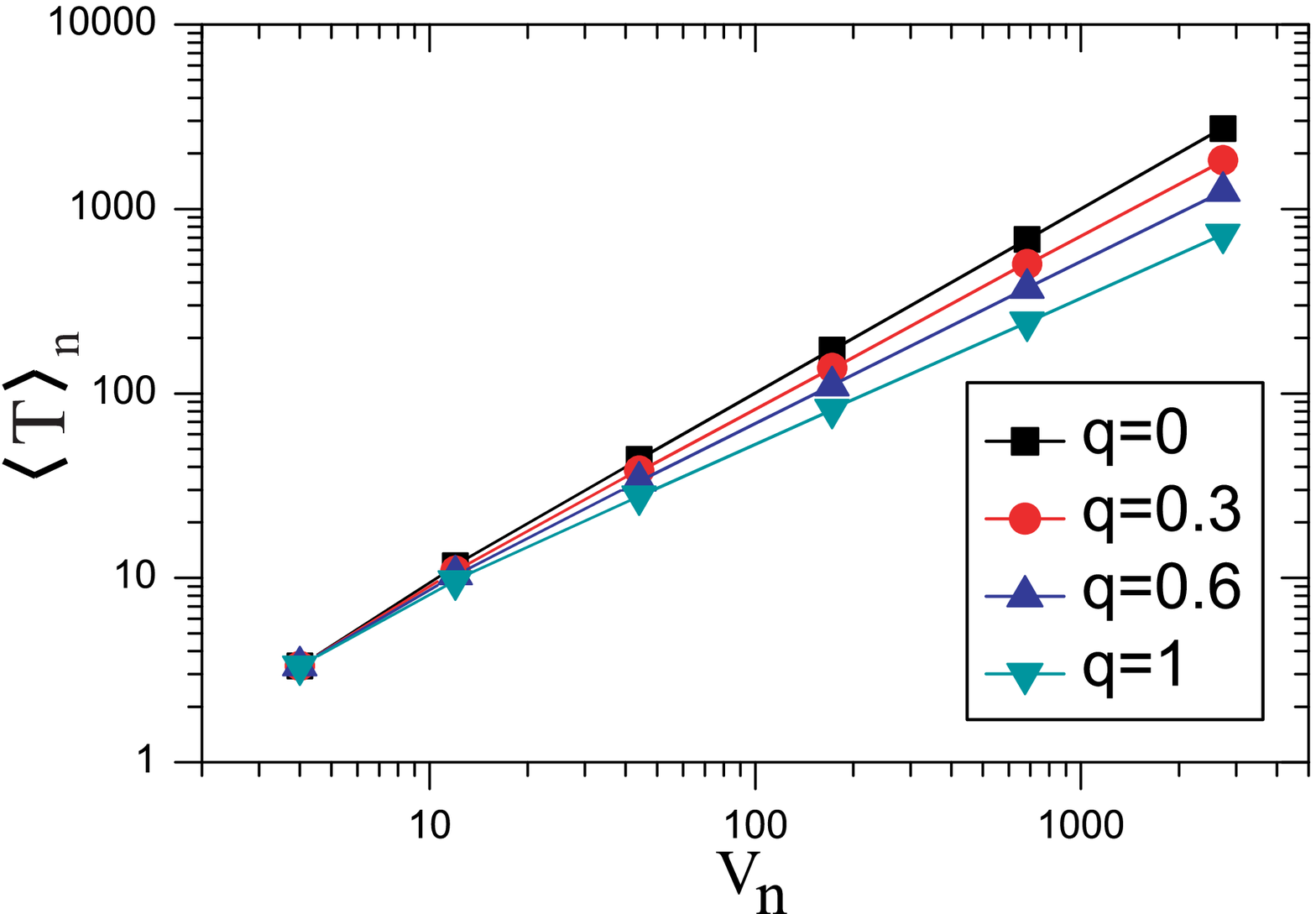}
\end{center}
\caption[kurzform]{\label{Time02} (Color online) Mean first-passage
time $\langle T \rangle_n$ versus the network order $V_n$ on a
log-log scale for various $n$ and $q$. The solid lines serve as
guides to the eye.}
\end{figure}

In order to obtain the dependence relation of MFPT with the network
order for $0 < q < 1$, we have performed extensive numerical
simulations for various networks with different iteration $n$ ($1
\leq n \leq 6$) and $q$ between $0$ and $1$. Figure~\ref{Time02}
illustrates the variation of MFPT with network order $V_n$, showing
that for all $0 \leq q \leq 1$, the MFPT grows as a power-law
function of $V_n$ with the exponent $\theta(q)$ changing with $q$:
When $q$ increases from 0 to 1, the exponent $\theta(q)$ decreases
from 1 to $\frac{\ln 3}{\ln 4}$.

\begin{figure}
\begin{center}
\includegraphics[width=.55\linewidth,trim=100 20 100 0]{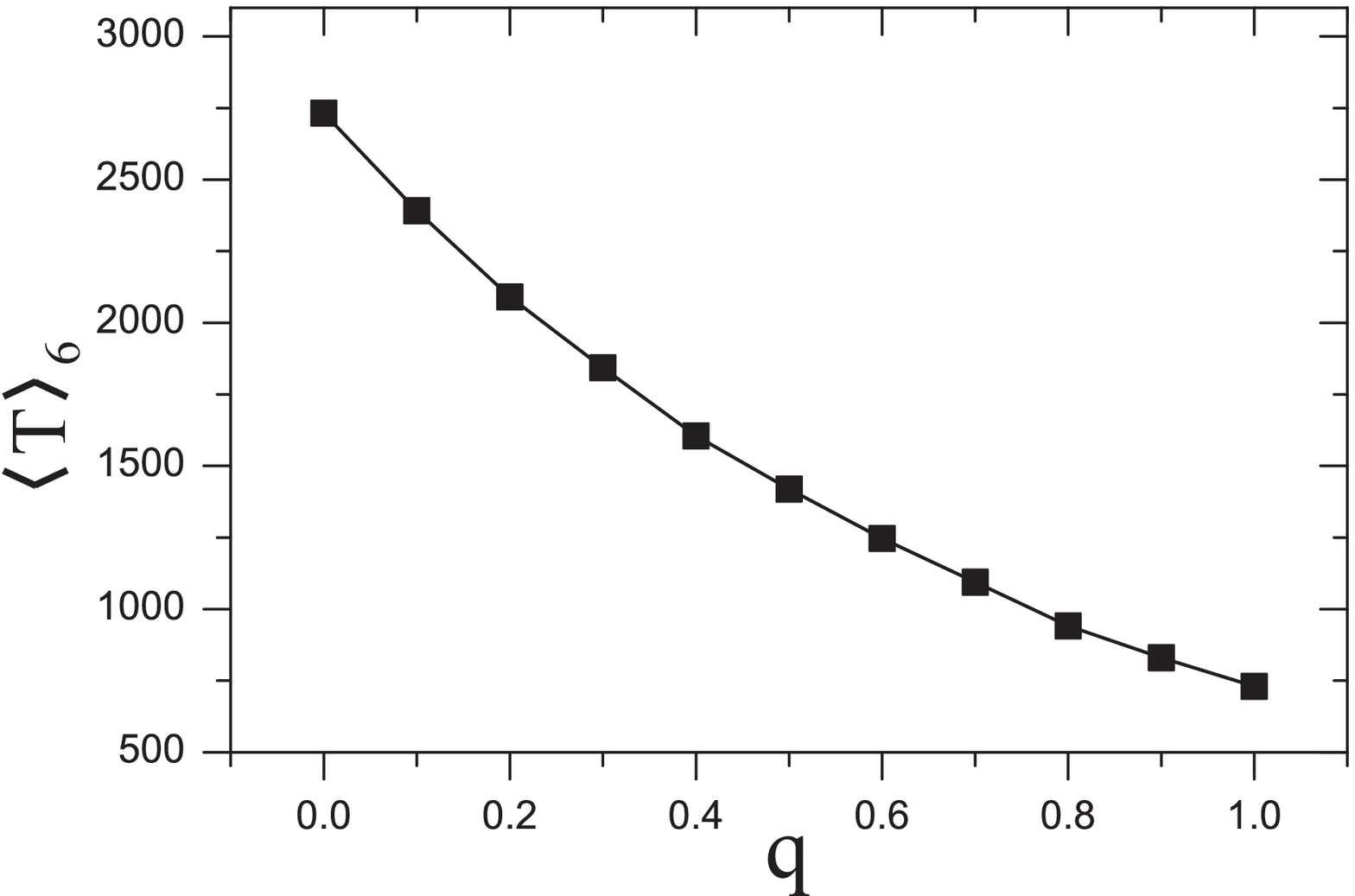}
\end{center}
\caption[kurzform]{\label{MFPTq} Dependence relation of MFPT on
parameter $q$.}
\end{figure}

From Fig.~\ref{Time02} we also know that the efficiency of trapping
process is closely related to parameter $q$: the larger the
parameter $q$, the higher of the efficiency of the trapping problem.
To show this concretely, we performed numerical calculation for
network $H_6$ with order $2732$ for different $q$. For each fraction
$q$ ($0 < q < 1$), all results are obtained by applying
Eq.~(\ref{MFPT5}) to an ensemble of 100 network realizations. In
Fig.~\ref {MFPTq}, we plot the MFPT, $\langle T\rangle_6$, as a
function of $q$. It is easily observed that when $q$ increases from
0 to 1, the MFPT decreases monotonically with $q$.

\section{Conclusions}

In summary, we have investigated the trapping issue on a family of
scale-free networks with identical degree sequence thus the same
degree distribution, which is controlled by a parameter $q$ ($0 \leq
q \leq 1$). We computed analytically or numerically the mean
first-passage time (MFPT) for the trapping problem on the networks
for various $q$. The obtained results show that for all $q$, the
MFPT grows as a power-law function of network order with the
exponent $\theta(q)$ dependent on $q$: when the parameter $q$ grows
from 0 to 1, the exponent $\theta(q)$ decreases from 1 to $\frac{\ln
3}{\ln 4}$, which indicates that power-law degree distribution alone
is not enough to characterize the trapping problem performed on
scale-free networks. Therefore, when one makes general statements
about the behavior of trapping issue on scale-free networks, care
should be needed.

\subsection*{Acknowledgment}

We thank Xing Li and Yichao Zhang for their assistance. This
research was supported by the National Basic Research Program of
China under Grant No. 2007CB310806; the National Natural Science
Foundation of China under Grants No. 60704044, No. 60873040, and No.
60873070; the Shanghai Leading Academic Discipline Project No. B114,
and the Program for New Century Excellent Talents in University of
China (Grant No. NCET-06-0376). W L Xie also acknowledges the
support provided by Hui-Chun Chin and Tsung-Dao Lee Chinese
Undergraduate Research Endowment (CURE).

\end{document}